\pdfoutput=1

\documentclass[aps,prl,reprint,twocolumn,groupedaddress,showpacs]{revtex4}

\usepackage{amsmath, amsfonts, amssymb, graphicx,cancel}

\newcommand{\ket}[1]{| #1 \rangle}
\newcommand{\braket}[2]{{\langle  #1 | #2 \rangle}}
\newcommand{\braketop}[3]{\langle #1 | #2 | #3 \rangle}

\newcommand{\dg}[0]{\dagger}

\newcommand{\vk}[0]{{\bf k}}

\newcommand{\vQ}[0]{{\bf Q}}

\newcommand{\vR}[0]{{\bf R}}

\renewcommand{\vr}[0]{{\bf r}}

\newcommand{\tr}[0]{{\rm Tr}}
\renewcommand{\epsilon}{\varepsilon}

\newcommand{\footnoteremember}[2]{
\footnote{#2}
\newcounter{#1}
\setcounter{#1}{\value{footnote}}
}
\newcommand{\footnoterecall}[1]{
\footnotemark[\value{#1}]
}

\begin{document}
\title{Microscopic theory of resonant soft x-ray scattering  in systems with charge order}

\author{David Benjamin}
\affiliation{Physics Department, Harvard University, Cambridge, Massachusetts 02138, USA
}

\author{Dmitry Abanin}
\affiliation{Physics Department, Harvard University, Cambridge, Massachusetts 02138, USA
}

\author{Peter Abbamonte}
\affiliation{Department of Physics and Frederick Seitz Materials Research Laboratory, University of Illinois, Urbana, IL 61801, USA}

\author{Eugene Demler}
\affiliation{Physics Department, Harvard University, Cambridge, Massachusetts 02138, USA
}

\date{\today}

\begin{abstract}
We present a microscopic theory of resonant soft x-ray scattering (RSXS) that accounts for the delocalized character of valence electrons.  Unlike past approaches defined in terms of form factors for atoms or clusters, we develop a functional determinant method that allows us to treat realistic band structures. This method builds upon earlier theoretical work in mesoscopic physics and accounts for both excitonic effects as well as the orthogonality catastrophe arising from interaction between the core hole and the valence band electrons. Comparing to RSXS measurements from stripe-ordered La$_{1.875}$Ba$_{0.125}$CuO$_4$, we show that the two-peak structure observed near the O {\it K} edge can be understood as arising from dynamic nesting within the canonical cuprate band structure.  Our results provide evidence for reasonably well-defined, high-energy quasiparticles in cuprates, and establishes RSXS as a bulk-sensitive probe of the electron quasiparticles.
\end{abstract}


\maketitle

{\it Introduction.}--\label{Introduction}
Resonant soft x-ray scattering (RSXS) is a powerful technique for exploring strongly-correlated quantum materials~\cite{Abbamonte2002a,Wilkins2003c,Abbamonte2005a}.  While neutron and non-resonant x-ray scattering cross sections are dominated by the contributions of nuclei and core electrons, RSXS couples selectively to valence electrons and provides an enormously enhanced sensitivity~\footnote{For example, in the case of the oxygen {\it K} edge in LBCO and LSCO the cross section for charge carriers is four orders of magnitude greater at resonance than away from resonance~\cite{Smadici2009a}.} to many-body correlations~\cite{Abbamonte2002a, Wilkins2003c, Dhesi2004, Grenier2004a, Abbamonte2004, Zegkinoglou2005, Schussler-Langeheine2005a, Abbamonte2005a,Fink2009b, Herrero-Martin2006a,Nazarenko2006a, Smadici2007b,Smadici2007c,Smadici2009a, Ghiringhelli2012, Hatton2005a}.  Further promise of RSXS comes from its ability to study a wide class of materials, including those available only in small samples and those with buried interfaces ~\cite{Smadici2007b,Smadici2007c,Smadici2009a}.  RSXS has recently been used to observe orbital order in manganites~\cite{Wilkins2003c, Dhesi2004, Grenier2004a} and ruthenates~\cite{Zegkinoglou2005}, hole crystallization in spin ladders~\cite{Abbamonte2004}, and charge order in cuprates~\cite{Abbamonte2005a,Fink2009b,Ghiringhelli2012}, nickelates~\cite{Schussler-Langeheine2005a}, and manganites~\cite{Herrero-Martin2006a,Nazarenko2006a}.

Although qualitative interpretation of RSXS data has already provided valuable insight into a variety of strongly correlated materials, a complete quantitative understanding
of these experiments is still lacking.  Up to now, most efforts to interpret resonant x-ray diffraction (RXD) experiments have adapted the use of atomic form factors from x-ray crystallography~\cite{Abbamonte2002a, Abbamonte2005a, Herrero-Martin2006a, Fink2009b}.  The form factor concept assumes optical locality, which is valid for ordinary x-ray diffraction, but breaks down in the resonant case if valence states are delocalized.  Some authors have attempted to account for this nonlocality by defining the form factor in terms of a cluster, rather than a single atom~\cite{Schussler-Langeheine2005a, Dhesi2004}.  But even this approach should break down if valence states are propagating quasiparticles.  Recent work of Abbamonte et al.~\cite{Abbamonte2011} showed that neglecting the finite lifetime of core holes and interaction of valence electrons with core holes allows one to relate RSXS spectra to the local electron Green's function measured in STM.  However, these neglected effects are expected to play an important role and it is not clear how accurately such a simplified analysis can explain RSXS spectra in real materials.  An approach based on the Bethe-Salpeter equation~\cite{Rohlfing1998} captures excitonic effects of the core hole but ignores the full many-body character of the core hole-Fermi sea interaction, including the orthogonality catastrophe.  The state of affairs in RSXS should be contrasted to the more established probes of valence electrons, ARPES~\cite{Damascelli2003, Lu2012,Campuzano2004} and STM~\cite{Fischer2007, Hoffman2002, McElroy2005, Vershinin2004, Howald2003, Yeh2009}, where it is often possible to read off spectral functions directly from measurements, facilitating the comparison of theoretical models with experimental results.

\begin{figure}
\begin{center}
\includegraphics[width=0.8\linewidth,natwidth=463,natheight=249]{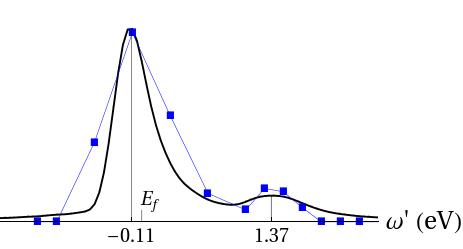}
\end{center}
\caption{Calculated RSXS spectrum for period-4 charge order for the canonical cuprate band-structure (parameters given below Eq. (\ref{H0}))
and core hole potential $U_0=-250$ meV. Horizontal axis shows shifted energy $\omega^\prime$ of scattered photons, where $\omega^\prime=0$ is the energy required to excite a core electron to $E_f$ in the absence of a core hole potential, vertical axis is intensity
of elastic scattering (arbitrary units). Squares give experimental data for LBCO from Ref.~\cite{Abbamonte2005a}.  Note that the position of the first peak is not $E_f$.  It is determined by both the dynamic nesting and the core hole potential~[60].}
\label{MainFigure}
\end{figure}

\begin{figure}
\begin{center}
\includegraphics[width=0.65\linewidth,natwidth=422,natheight=460]{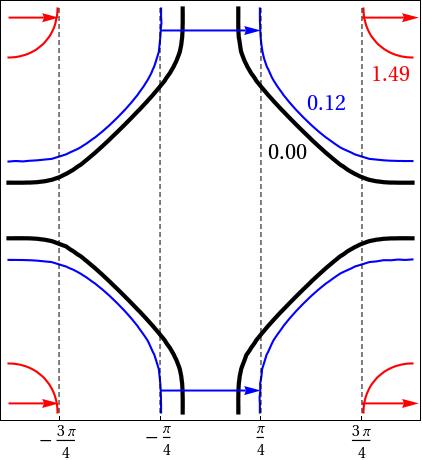}
\end{center}
\caption{Dynamic nesting in the cuprate band structure.  Nested segments of the $E=E_f+0.12$ eV and $E=E_f+1.49$ eV contours are shown in blue and red.  The Fermi surface is shown in black.  The lines $k_x = -3\pi/4,-\pi/4,\pi,4,3\pi/4$ (dashed) are a visual guide.}
\label{DynamicNesting}
\end{figure}

In this paper we present the first microscopic model of elastic RSXS in systems with charge order in the valence band, such as striped  high-T$_{\rm c}$ cuprates~\cite{Tranquada1995c, White2000, Hoffman2002, Kivelson2003a, Millis2007, Vojta2009b, Berg2009a, Zaanen1989}.  We develop a theoretical approach that allows us to analyze RSXS spectra in the case of itinerant valence electrons and include realistic bandstructures. Our formalism accounts for excitonic effects and the orthogonality catastrophe arising from the interaction of valence electrons with core holes
(within the approximation of noninteracting valence electrons our analysis is exact) as well as the finite lifetime of core holes.  We show that the two-peak spectrum observed in experiments at the O \textit{K} edge of La$_{1.875}$Ba$_{0.125}$CuO$_4$~\cite{Abbamonte2005a,Fink2009b} can be explained by dynamical nesting of the ``standard" band structure of cuprates (see Figs.~\ref{MainFigure} and \ref{DynamicNesting}).  We find that interaction of valence electrons with the core hole changes the spectrum significantly.  For a physically reasonable core hole potential we obtain quantitative agreement with the experimental data on underdoped La$_{1.875}$Ba$_{0.125}$CuO$_4$  (LBCO) near the 1/8-anomaly~\cite{Abbamonte2005a} (similar spectra were observed for La$_{1.8-x}$Eu$_{0.2}$Sr$_{x}$CuO$_4$(LESCO) in \cite{Fink2009b}).                                   Our results suggest that RSXS at the O {\it K} edge can be directly connected to the known band structure, providing a new, bulk-sensitive probe of electron quasiparticles that is complementary to ARPES and STM techniques.

{\it Theoretical formalism for elastic RSXS}.-- Following Ref. \cite{Abbamonte2011} we consider an effective single
band model describing resonant absorption and emission of photons
\begin{eqnarray}
{\cal H}_{\rm int} = \sum_{j,\vk,\lambda} V(\vk,\lambda) \left( d_j^\dg c_j a_{\vk,\lambda} e^{ i \vk \cdot \vr} + {\rm h.c.} \right)
\end{eqnarray}
Here $c_j$ and $d_j$ are annihilation operators of electrons on site $j$ in the core orbital and valence band respectively,
$a_{\vk,\lambda}$ annihilates a photon with momentum $\vk$ and polarization $\hat{\epsilon}_{\vk,\lambda}$, and
 $V(\vk,\lambda)$ are matrix elements whose precise form is not important to us.
Resonant scattering is a second order process in ${\cal H}_{\rm int}$
\begin{eqnarray}
\label{intensity}
I(\omega) = \sum_f \left| \sum_n \frac{ \braketop{f}{{\cal H}_{\rm int}}{n} \braketop{n}{{\cal H}_{\rm int}}{i}}
{E^N_i -\tilde{E}^{N+1}_n+ \omega + i \Gamma/2} \right|^2
\end{eqnarray}
Here $\ket{i} $ is the initial state of the system with $N$ electrons in the valence band, no core holes, and one incoming photon with momentum $\vk_i$; $\ket{n}$ is the intermediate state
with one core hole and one extra electron in the valence band, and no photons; $\ket{f}$
is the final state with the core levels filled again, $N$ electrons in the valence band
and one outgoing photon with momentum $\vk_f$. We also introduced $E^N_i$ and $\tilde{E}^{N+1}_n$
to denote energies of the initial and intermediate electron states respectively. Note that the latter includes the
potential of the core hole. $\Gamma$ is the decay rate of the core hole, including radiative and non-radiative processes.  We focus on elastic scattering by ordering wavevector $\vQ$ where we can take $\ket{f} = \ket{i}$ and $\vk_f = \vk_i + \vQ$.

Because the core hole is immobile it must be re-filled on the same site $j$ on which it is created and it may be subsumed into a static potential that acts on the valence electrons of the intermediate state.  The core hole contributes the trivial matrix element $\braketop{1}{c^\dg_j}{0}\braketop{0}{c_j}{1}=1$.  Therefore, Eq.~(\ref{intensity}) takes the following form:

\begin{align}
I(\omega,\vQ) &\propto
	\left| \sum_{j,n,\sigma} e^{-i \vQ \cdot \vr_j} \frac{\braketop{i}{d_{j\sigma}}{n}\braketop{n}{d^\dg_{j\sigma}}{i}}{E_i-\tilde{E}^{N+1}_n+\omega +i\Gamma/2} \right|^2
	\label{EnergyDomain} \\
	&=
	\left| \sum_{j\sigma} e^{-i \vQ \cdot \vr_j} \int_0^\infty  e^{-(i\omega+\Gamma/2) t} S_{j\sigma}(t) dt \right|^2,
	\label{TimeDomain}
\end{align}
where $S_{j\sigma}(t) = \braketop{i}{d_{j\sigma}  e^{-i {\cal H}_1(j) t} d^\dg_{j\sigma} e^{-i {\cal H}_0 t}}{i}$ and ${\cal H}_{0,1}$ is the Hamiltonian of the valence electrons with and without the core hole potential. Here the matrix elements refer to the valence electron Fock space.  When the core hole potential is zero, ${\cal H}_1(j)={\cal H}_0$ and $S_{j\sigma}(t)$ reduces essentially to the retarded Green's function\cite{Abbamonte2011}.  If we measure photon frequencies relative to the difference between the chemical potential of the valence band $\mu$
and the energy of the core state electron $\xi_c$, $\omega' = \omega - (\mu-\xi_c)$, we can set $E_i=0$ so that ${\cal H}_0\ket{i}=0$.  The intermediate states $\ket{n}$ are eigenstates of
\begin{eqnarray}
{\cal H}_1(j) = {\cal H}_0 + U(\vr - \vr_j) ,
\label{H1}
\end{eqnarray}
where $U$ is the potential due to the core hole at site $j$ and ${\cal H}_0 $ is given in the grand canonical ensemble.
Equation (\ref{TimeDomain}) also applies to thermal ensemble at temperature $T=1/\beta$, provided that we use
\begin{equation}
S_{j\sigma}(t) =
\frac{\tr_F \left[ d_{j\sigma}  e^{-i {\cal H}_1(j) t} d^\dg_{j\sigma} e^{-\beta {\cal H}_0} \right]}{\tr_F \left[e^{-\beta {\cal H}_0} \right]}.
\label{FockTrace}
\end{equation}
The subscript $F$ on the trace denotes that it is taken with respect to Fock space.

Eqs. (\ref{TimeDomain} - \ref{FockTrace}) are very general and apply
for an arbitrary interacting valence band Hamiltonian ${\cal H}_0 $.  While Eq.~(\ref{FockTrace}) was formulated for the single band case, it can be generalized to the multi orbital case by adding orbital indices to electron operators and restoring orbital-dependent matrix elements for absorption and emission.  In the remainder of this paper we will limit our discussion to the model of non-interacting electrons.
This simplifying assumption is justified as long as the lifetime of electron states
in the valence band is longer than the lifetime $\Gamma^{-1}$ of the core hole (see also discussion below).

The trace in Eq.~(\ref{FockTrace}) is reminiscent of the orthogonality catastrophe of x-ray absorption~\cite{Mahan1967,Nozieres1969} and the Fermi edge singularity of mesoscopic transport~\cite{Muzykantskii2003, Abanin2005}.  It expresses the many-body overlap of the initial Fermi sea with the perturbed Fermi sea that time evolves under ${\cal H}_1$ and the single-particle dynamics of the extra electron injected into the valence band at site $j$.  It can be reduced to a product of determinants and inverse matrix elements of operators in single-particle Hilbert space~\cite{Combescot1971,Klich2004}.  As an example, by appealing to a basis in which ${\cal H}_0$ is diagonal, ${\cal H}_0 \ket{\phi_\alpha} = \xi_\alpha \ket{\phi_\alpha}$, the trace $\tr_{F,\sigma} \left[e^{-\beta {\cal H}_0} \right]$ factors as $\prod_\alpha \sum_{n_\alpha=0,1} e^{-\beta n_\alpha \xi_\alpha}=\prod_\alpha (1+e^{-\beta \xi_\alpha})$.  This in turn is the product of eigenvalues of the {\it single-particle} operator $1+e^{-\beta H_0}$, hence it equals $\det (1+e^{-\beta H_0})$.  Here
$H_{0,1}$ refer to single particle operators for a single spin component of electrons.  What permits further progress is that by the Baker-Campbell-Haussdorff Lemma $e^{-i {\cal H}_1 t} e^{-\beta {\cal H}_0}$ can be written as $e^W$ where $W$ is quadratic, and hence the remaining traces can be evaluated in the same spirit~\cite{Klich2004}.  We obtain\footnoteremember{supp}{See Supplemental Material for derivation of trace formulae, variation of strength and form of the core hole potential, variation of hopping strength, and comments on slightly-inelastic scattering.}
\begin{equation}
S_j(t) =
	 \det \left( (1-\hat{N}) + \hat{U}_j(t) \hat{N} \right)^2 \left(\frac{\hat{N}}{1-\hat{N}} + \hat{U}_j^{-1}(t) \right)^{-1}_{jj},
	\label{MainResult}
\end{equation}
where $\hat{N} \equiv (1+\exp(\beta \hat{H}_0))^{-1}$ is the occupation number operator and $\hat{U}_j(t) \equiv e^{-i \hat{H}_1(j) t}$ is the intermediate state time evolution operator.  The determinantal factor corresponds to the many-particle dynamics of the Fermi sea.  It is squared because the orthogonality catastrophe occurs for each of the two spin species regardless of the spin of the photo-excited electron\footnoterecall{supp}\!\!. One can interpret it as follows: the argument is an operator that time-evolves only those states that are initially occupied, and taking the determinant computes an overlap of Slater states.  The matrix element part corresponds to single-particle dynamics of the intermediate photoelectron.  It is a local Green's function for propagation of a single electron from site $j$ to site $j$, modified  by the Pauli-blocking term $N/(1-N)$.  For period-$p$ order, one needs to sum over $p$ inequivalent sites $j$.  The determinant can be evaluated efficiently for a finite system, converging by a system size of $25 \times 25$.  Eqs.~(\ref{TimeDomain}) and (\ref{MainResult}) constitute a convenient formula for calculating RSXS spectra in the approximation of noninteracting electrons. They treat exactly the interaction of electrons with the core hole and finite lifetime of the core hole.

\label{Cuprates}
{\it RSXS of cuprates.}--
We apply Eqs.~(\ref{TimeDomain}) and (\ref{MainResult}) to charge order in an effective one-band model of the cuprates
\begin{eqnarray}
H_0 = \sum_\vk \xi_\vk d^\dg_{\vk} d_{\vk}+V \sum_\vk \left(d^\dg_{\vk+\vQ} d_\vk + d^\dg_\vk d_{\vk+\vQ} \right).
\label{H0}
\end{eqnarray}
Eq.~(\ref{H0}) is a mean-field phenomenological description of charge ordering~\cite{Gruner1994, Podolsky2003, Kivelson2003a, Hackl2010a, Vojta2009b, Sedrakyan2010, Yao2011, Balatsky} that applies regardless of its microscopic origin. Possible mechanisms include electron-electron interactions, in
which case charge order is often called stripes~\cite{Zaanen1989, Schulz1990}, and nesting of the Fermi surface and electron-phonon interactions~\cite{Li2006, Shen2005, Andersen1991}.

We use the tight-binding dispersion $\xi_\vk = -\sum_\vr e^{i \vk \cdot \vr} t_\vr - \mu$
and parameters $t_{(1,0)}=340, \, t_{(1,1)} = -32, \, t_{(2,0)} = 25, \, t_{(2,1)} = 31$ meV
 characteristic of LBCO~\cite{Markiewicz2005}.  For simplicity we ignore $k_z$ dispersion, which would at most smear energy peaks by an amount $t_z \lesssim $ 50 meV~\cite{Markiewicz2005}.  Fig.~\ref{MainFigure} presents an RSXS spectrum for a contact core hole potential $U(\vr-\vr_j)=U_0 \delta_{\vr,\vr_j}$.  We have also calculated spectra using Yukawa potentials of various ranges and found similar results\footnoterecall{supp}\!.  We have chosen a realistic core hole lifetime $\Gamma = 250$ meV.

{\it Two peak structure}. Fig.~\ref{MainFigure} shows the calculated intensity of RSXS as a function of initial photon energy.  The two peak structure of the spectrum agrees well with experimental findings~\cite{Abbamonte2005a,Fink2009b}.
A simple physical argument shows that the two peak structure is a robust feature of the cuprate band structure.  Consider the simplifying limit of zero core hole potential.
In this case the intermediate eigenstates are obtained by adding an electron in some eigenstate $\ket{\phi}$ of $H_0$ to the Fermi sea, and the energy domain expression Eq.~(\ref{EnergyDomain}) reduces to
\begin{equation}
I(\omega,\vQ) \propto
	\left| \sum_{j,\phi} e^{-i \vQ \cdot \vR_j}\frac{(1-n_F(E_\phi))\left|\braket{\phi}{j}\right|^2}{E_i-E_\phi+\omega +i\Gamma/2}  \right|^2.
\end{equation}
Unoccupied states $\phi$ contribute strongly when their energy is in resonance and the Fourier transform of their density at wavevector $\vQ$ is large.  In the presence of a CDW potential, Bloch states $\ket{\vk}$ and $\ket{\vk+\vQ}$ hybridize to form eigenstates with non-trivial density.  This occurs most readily when $\xi_\vk, \, \xi_{\vk+\vQ}$ are nearly degenerate.  The RSXS intensity at energy $E$ comes from points on the surface of constant energy $E$ that are separated by wavevector $\vQ$~\cite{Podolsky2003}.  Usually these are isolated pairs of points, but at certain energies the surfaces are nested so that sections separated by $\vQ$ move in parallel (see Fig.~\ref{DynamicNesting}).  At these energies there is a large density of hybridized states.  This is the phenomenon of dynamic nesting.
While Fermi surface nesting, which is just dynamic nesting at $E=E_f$, is uncommon, dynamic nesting is a generic consequence of symmetry.  Consider the two-dimensional cuprate Brillouin zone and period-4 CDW wavevector $\vQ=(\pi/2,0)$.  Any Bloch state $\ket{\vk}$ on the lines $k_x = -\pi/4$ and $k_x = 3 \pi/4$ is degenerate with  $\ket{\vk+\vQ}$.  Constant energy contours with segments tangent to the line $k_x = -\pi/4 \, (-3 \pi / 4)$ also have segments tangent to $k_x = \pi/4 \, (3 \pi / 4)$; these symmetry-equivalent segments are dynamically nested.  Fermi surface nesting is not generic because there is no particular reason why the contour $E=E_f$ should be tangent to the lines (or, in three dimensions, a plane).  Dynamic nesting, on the other hand, occurs when \textit{some} energy contour is tangent to the lines.  For our choice of LBCO hopping strengths the energy contours exhibiting dynamic nesting correspond to energies 0.1 eV and 1.5 eV above the Fermi level-- which are separated by nearly the same amount as RSXS peaks.  The small discrepancy is due to the tendency of the core hole potential to widen the distance between peaks and vanishes when we set $U_0 = 0$\footnoterecall{supp}.

A simplified model neglecting the core hole potential explains the spectrum and its two peaks qualitatively but does not give the correct relative weights of the two peaks\footnoterecall{supp}\!\!. Including the core hole potential yields quantitative agreement with experiments. The core hole potential has a weak effect on the energy separation between the two peaks but dramatically decreases the intensity of the high-energy peak\footnoterecall{supp}\!\!.
A core hole potential strength $U_0=-250$ meV, which is reasonable for a screened core
hole interacting with valence electrons, reproduces the experimental ratios of peak intensities.
A non-physical repulsive core hole decreases the intensity of the low-energy peak.  The discussion of Ref.~\cite{Abbamonte2011} connecting the RSXS spectrum to the electron spectral function thus remains largely accurate in the presence of a weak core hole potential.  However, strong core hole potentials yield spectra with qualitative features, such as a lack of a high-energy peak, that give misleading conclusions in analyses based only on the spectral function.  For example, we attribute the absence of a second peak in RSXS at the Cu {\it L}$_{3/2}$ edge~\cite{Abbamonte2005a} to a strong Cu core hole potential.
The spectrum is robust to changes in the core hole lifetime $\Gamma$, which broadens the peaks, and the CDW
strength $V$, which scales the overall intensity.   Small changes in the hopping strengths have little effect;
multiplying them by a uniform factor affects the distance between peaks.

Our calculations provide the first quantitative explanation for the two peak structure observed in LBCO and LESCO
~\cite{Abbamonte2005a,Fink2009b}. An earlier interpretation of the two peaks as arising from the lower and upper Hubbard bands, the so-called ``spatially-modulated Mottness", was not supported by quantitative analysis. Moreover, a separation of $\sim$ 1.9 eV between peaks is found in x-ray absorption spectroscopy (XAS) of LBCO and LSCO~\cite{Peets2009, Abbamonte2005a}.  According to the lower/upper Hubbard band interpretation, in which between peaks there is a gap, the separation between peaks in RSXS must be at least as large as the separation in XAS.  Thus we think that dynamical nesting of the band structure provides a more natural interpretation of the two peak structure observed in LBCO and LESCO.

{\it Discussion}.-- We now comment on the specific values of the band structure that we used in our analysis.  Ab-initio LDA calculations on LSCO give $t_{(1,0)}=430, \, t_{(1,1)} = -40, \, t_{(2,0)} = 30, \, t_{(2,1)} = 35$ meV~\cite{Markiewicz2005} while fitting of the ARPES spectra gives $t_{(1,0)}=250, \, t_{(1,1)} = -25, \, t_{(2,0)} = 20, \, t_{(2,1)} = 28$~\cite{Markiewicz2005}.  The ratios among tight-binding parameters are nearly identical for both cases, so the band structure is well-known up to an overall truncation factor.  The two peak character of the RSXS spectra appears for both band structures with nearly the same relative intensities of the two peaks. We find that taking either the LDA or ARPES dispersions gives peaks separated by 1.7 and 1.3 eV. We obtain the best fit to RSXS data by choosing parameters halfway between the two. We point out that it is not so surprising that
band structure obtained from the ARPES data does not provide the best agreement with the RSXS spectra.  ARPES data only exist within 200 meV of the Fermi surface~\cite{Ino2002}, where the renormalization effect due to interactions is strongest, while we are interested in features at much higher energy.  Additionally, it has been suggested that ARPES tends to underestimate electron dispersion relative to x-ray experiments~\cite{StefanHufner,Markiewicz2005}. Another important issue is our approximation of  non-interacting electrons. The key quantity entering our analysis
is a generalized propagator (\ref{FockTrace}). The effect of many-body correlations is to introduce decay of an electron into other excitations, but as far as the Green's function is concerned this simply contributes an imaginary part to the electron's self-energy (we assume that the effective one-band model we use has already incorporated renormalization via the {\it real} part of self-energies).  Furthermore, if the decay of the electron is slow compared to the decay of the core hole, any broadening introduced by electron interactions will be hidden within the width $\Gamma$.  Conversely, if the electron decays very rapidly, RSXS peaks will be broadened into oblivion.  Therefore, the presence of peaks in an RSXS spectrum puts an upper bound on the imaginary self-energy and implies that excitations resemble well-defined quasiparticles.  We note that recent DMFT calculations~\footnote{A. Georges, unpublished} have found long-lived electron quasiparticles in the Hubbard model well above the Fermi energy, in contrast to short-lived hole-like excitations.  RSXS, which probes high-energy  electron excitations, complements ARPES, which probes hole-like excitations, and magnetic oscillation experiments~\cite{LeBoeuf2007a, Doiron-Leyraud2007a, Vignolle2011b}, which probes only excitations near the Fermi energy.

{\it Outlook}.--The predictions of our model can be checked in future experiments.  For example, recent work on charge order in underdoped YBCO~\cite{Ghiringhelli2012}, which was performed at energies corresponding to Cu {\it L} edges, could be repeated at the O {\it K} edge.  We expect, as in LBCO, two peaks at energies determined by band structure.  Also, systems with checkerboard charge order, with coexisting Fourier components $\vQ_x$ and $\vQ_y$, will exhibit a harmonic at $\vQ_x + \vQ_y$.  If the latter harmonic is sufficiently strong, an RSXS signal will appear at this wavevector. One can see that this ordering wavevector also has dynamic nesting at two energies, so we expect to find a two peak spectrum\footnote{D. Benjamin, unpublished}.

{\it Summary}.--
We have developed a microscopic model of RSXS that takes into account the itinerant character of valence electrons and excitonic effects. We showed that a simple physical picture of
dynamical nesting found in the canonical band structure of cuprates gives rise to a two peak structure, while the core hole potential is necessary for quantitative agreement with the data.  Our analysis shows that even at high energies electronic excitations behave like sufficiently well-defined quasiparticles described by the canonical band structure.

{\it Acknowledgements}.--
We thank A. Georges and J. Sau for useful discussions.  This work was supported by Harvard-MIT CUA,
NSF Grant No. DMR-07-05472,  and the ARO-MURI on Atomtronics (DB, DA, ED); and the U.S. Department of Energy grant DE-FG02-06ER46285 (PA).

\bibliography{library}

\end{document}


\title{Supplementary Materials}

\author{David Benjamin}
\affiliation{Harvard University Department of Physics}

\author{Dmitry Abanin}
\affiliation{Harvard University Department of Physics}

\author{Eugene Demler}
\affiliation{Harvard University Department of Physics}

\date{\today}

\maketitle

\section{Core Hole Matrix Elements}
Here we use the method of Klich to compute expressions of the form
\begin{equation}
A = \frac{\tr_F \left[ d_{j,\sigma}  e^{-i{\cal H}_1 t} d^\dg_{j,\sigma} e^{-\beta {\cal H}_0} \right]}{\tr_F \left[e^{-\beta {\cal H}_0} \right]},
\label{OriginalTrace}
\end{equation}
where ${\cal H}_{0,1}$ are quadratic Hamiltonians and $d_j$ annihilates the single-particle Wannier orbital $\ket{j}$.   We write $\tr_F$ and $\tr_H$ for traces with respect to many-particle Fock space and single-particle Hilbert space, respectively.We split them into commuting parts that act on electrons of spins $\sigma$ and $\bar{\sigma}$: ${\cal H}_{0(1)} = {\cal H}_{0(1),\sigma}+ {\cal H}_{0(1),\bar{\sigma}}$.  Then the traces factorize into different spin sectors:
\begin{equation}
A = \frac{\tr_F \left[ d_{j,\sigma}  e^{-i{\cal H}_{1,\sigma} t} d^\dg_{j,\sigma} e^{-\beta {\cal H}_{0,\sigma}} \right]}{\tr_F \left[e^{-\beta {\cal H}_{0,\sigma}} \right]}
\frac{\tr_F \left[ e^{-i{\cal H}_{1,\bar{\sigma}} t}  e^{-\beta {\cal H}_{0,\bar{\sigma}}} \right]}{\tr_F \left[e^{-\beta {\cal H}_{0,\bar{\sigma}}} \right]},
\label{OriginalTraceSplitSpin}
\end{equation}
Each trace is spin-independent, so let us introduce
\begin{align}
X_0 =& -\beta{\cal H}_{0,\sigma} = -\beta{\cal H}_{0,\bar{\sigma}} \\
X_1 =& -it{\cal H}_{1,\sigma} = -it{\cal H}_{1,\bar{\sigma}},
\end{align}
where equality is up to isomorphism of spin-$\sigma$ and spin-$\bar{\sigma}$ subspaces.  That is, $X_{0,1}$ act on the Fock space of a spinless conduction band.  Equation (\ref{OriginalTraceSplitSpin}) then reads
\begin{equation}
A = \frac{\tr_F \left[ d_{j}  e^{X_1} d^\dg_{j} e^{X_0} \right]}{\tr_F \left[e^{X_0} \right]}
\frac{\tr_F \left[ e^{X_1}  e^{X_0} \right]}{\tr_F \left[e^{X_0} \right]},
\label{OriginalTraceSplitSpinX}
\end{equation}

Define $X_2$ such that $\exp(X_1)\exp(X_0) \equiv \exp(X_2)$.  By the Baker-Campbell-Haussdorff Lemma, $X_2$ is quadratic.  Label the eigenstates of $X_i$ as follows:
\begin{equation}
X_i \ket{\alpha,i}=\omega_{\alpha,i} \ket{\alpha,i}, \, X_i= \sum \omega_{\alpha,i} d^\dg_{\alpha,i} d_{\alpha,i} = \sum \omega_{\alpha,i} n_{\alpha,i}
\end{equation}
Change of basis is done with $d_j = \sum_\alpha \braket{j}{\alpha,i} d_{\alpha,i}$ etc, and the useful corollary
\begin{equation}
\sum_\alpha \ket{\alpha,i} d_{\alpha,i} = \sum_{\alpha} \ket{\alpha,i}\bra{\alpha,i} \left( \sum_\beta \ket{\beta,j} d_{\beta,j} \right) =  \sum_\beta \ket{\beta,j} d_{\beta,j}.
\label{ChangeBasisCorollary}
\end{equation}

The denominator traces in Equation~\ref{OriginalTraceSplitSpinX} are the simplest to evaluate and are a useful warm-up for the numerator traces.  The first step in the basis $\{ \ket{\alpha,0} \}$ in which $X_0$ is diagonal and the many-body trace factorizes in terms of traces over the occupation numbers $n_{\alpha,0} = 0, 1$ of eigenstates:
\begin{equation}
\tr_F \left[e^{X_0} \right] = \tr_F \left[e^{\sum_\alpha \omega_{\alpha,0} n_{\alpha,0}} \right]=\prod_\alpha \sum_{n_{\alpha,0}=0,1} e^{\sum_\alpha \omega_{\alpha,0} n_{\alpha,0}} = \prod_\alpha (1+ e^{\omega_{\alpha,0}}).
\label{FirstDenominatorEquation}
\end{equation}
Next we use the fact that $(1+e^{\omega_\alpha,0})$ are the eigenvalues of the {\it single-particle} operator $1+ \exp(X_{0,{\rm sp}})$, where $X_{0,{\rm sp}}$ is the projection of $X_0$ into Hilbert space.  (We define $X_{1,{\rm sp}}$ and $X_{2,{\rm sp}}$ analogously).  Furthermore, the product of an operator's eigenvalues is its determinant, so we have
\begin{equation}
\tr_F \left[e^{X_0} \right]  = \prod_\alpha (1+ e^{\omega_{\alpha,0}}) = \det \left( 1+ e^{X_{0,{\rm sp}}}\right).
\label{DenominatorResult}
\end{equation}

Of the two distinct numerators in Equation~(\ref{OriginalTraceSplitSpinX}) the simpler one, which has no injected electron, follows almost immediately fromEquation~(\ref{DenominatorResult}):
\begin{equation}
\tr_F \left[e^{X_1}e^{X_0} \right]=\tr_F \left[e^{X_2} \right] = \det \left( 1+ e^{X_{2,{\rm sp}}}\right)= \det \left( 1+ e^{X_{1,{\rm sp}}}e^{X_{0,{\rm sp}}}\right)
\label{EasyNumeratorResult}
\end{equation}
Note that we never have to go through the exercise in commutator algebra to obtain $X_2$ explicitly.  All we needed was that it is quadratic.

To calculate the last and most difficult trace, we again want to write it as a product of traces over occupation numbers 0 and 1.  This means that we wish to put the $d$ and $d^\dg$ next to one another to form a number operator $n$.  We use the cyclic property of the trace to put $d_j$ to the right of $d^\dg_j$ and switch to a basis in which $X_0$ is diagonal:  :
\begin{equation}
\tr_F \left[ d_j  e^{X_1} d^\dg_j e^{X_0} \right]=
\tr_F \left[  e^{X_1} d^\dg_j e^{X_0} d_j\right]=
\sum_{\alpha,\alpha^\prime} \braket{j}{\alpha,0} \tr_F \left[  e^{X_1} d^\dg_{\alpha^\prime,0} e^{X_0} d_{\alpha,0} \right] \braket{\alpha^\prime,0}{j}.
\end{equation}
In this basis we may exploit the commutator $ d^\dg_{\alpha,0}e^{X_0} = e^{X_0} d^\dg_{\alpha,0} e^{-\omega_{\alpha,0}}$ to get
\begin{equation}
\tr_F \left[ d_j  e^{X_1} d^\dg_j e^{X_0} \right] = 
\sum_{\alpha,\alpha^\prime} \braket{j}{\alpha,0} \tr_F \left[  e^{X_1} e^{X_0} d^\dg_{\alpha^\prime,0}  d_{\alpha,0} \right] e^{-\omega_{\alpha^\prime,0}}\braket{\alpha^\prime,0}{j}
\label{AStep}
\end{equation}
Next we wish to combine $e^{X_1}$ and $e^{X_0}$ as $e^{X_2}$ and use 
Equation~(\ref{ChangeBasisCorollary}) with $i=0, \, j=2$ to switch to the eigenbasis of $X_2$.  We first need to remove the factor $e^{-\omega_{\alpha^\prime,0}}$ via
\begin{equation}
e^{-\omega_{\alpha^\prime,0}}\bra{\alpha^\prime,0} = 
\bra{\alpha^\prime,0}e^{-X_{0,{\rm sp}}}.
\label{ScalarToOperator}
\end{equation}
By Equations~(\ref{ChangeBasisCorollary}) and (\ref{ScalarToOperator}), Equation~(\ref{AStep}) becomes
\begin{align}
\tr_F \left[ d_j  e^{X_1} d^\dg_j e^{X_0} \right] =& 
\sum_{\beta,\beta^\prime} \braket{j}{\beta,2} \tr_F \left[  e^{X_2} d^\dg_{\beta^\prime,2}  d_{\beta,2} \right] \braketop{\beta^\prime,2}{e^{-X_{0,{\rm sp}}}}{j} \nonumber \\
=&\sum_{\beta} \braket{j}{\beta,2} \tr_F \left[  e^{\sum \omega_{\gamma,2} n_{\gamma,2}} n_{\beta,2} \right] \braketop{\beta,2}{e^{-X_{0,{\rm sp}}}}{j}.
\label{DeepInEquations}
\end{align}
We used the fact that $\beta\ne\beta^\prime$ is purely off-diagonal and has vanishing trace.  We evaluate the trace in Equation~(\ref{DeepInEquations}) the same way as Equation~(\ref{FirstDenominatorEquation}), with the slight complication that we must distinguish the case $\gamma=\beta$ and $\gamma \ne \beta$:
\begin{align}
\tr_F \left[  e^{\sum \omega_{\gamma,2} n_{\gamma,2}} n_{\beta,2} \right] =&
\sum_{n_{\beta,2}=0,1}e^{\omega_{\beta,2} n_{\beta,2}}n_{\beta,2} \prod_{\gamma \ne \beta} \sum_{n_{\gamma,2}=0,1}e^{\omega_{\gamma,2} n_{\gamma,2}}\nonumber \\
 =&e^{\omega_{\beta,2}}\prod_{\gamma \ne \beta} \left(1+e^{\omega_{\gamma,2}}\right) 
= \frac{e^{\omega_{\beta,2}}}{1+e^{\omega_{\beta,2}}}\prod_{\gamma} \left(1+e^{\omega_{\gamma,2}}\right)\nonumber \\
 =&
\frac{e^{\omega_{\beta,2}}}{1+e^{\omega_{\beta,2}}} \det \left(1+e^{X_{2,{\rm sp}}}\right).
\label{ComplicatedTraceResult}
\end{align}
When we substitute Equation~(\ref{ComplicatedTraceResult}) back into Equation~(\ref{DeepInEquations}), we again use the trick of replacing a scalar by an operator,
\begin{equation}
\frac{e^{\omega_{\beta,2}}}{1+e^{\omega_{\beta,2}}} \bra{\beta,2}=
\bra{\beta,2}\frac{ e^{X_{2,{\rm sp}}} }{1+e^{X_{2,{\rm sp}}}},
\end{equation}
to obtain
\begin{align}
\tr_F \left[ d_j  e^{X_1} d^\dg_j e^{X_0} \right] =& 
\det \left(1+e^{X_{2,{\rm sp}}}\right) \sum_{\beta} \braket{j}{\beta,2} \braketop{\beta,2}{\frac{ e^{X_{2,{\rm sp}}} }{1+e^{X_{2,{\rm sp}}}} e^{-X_{0,{\rm sp}}}}{j} \nonumber \\
=& \det \left(1+e^{X_{2,{\rm sp}}}\right)  \braketop{j}{\frac{ e^{X_{2,{\rm sp}}} }{1+e^{X_{2,{\rm sp}}}} e^{-X_{0,{\rm sp}}}}{j}
\label{ComplicatedNumeratorResult}
\end{align}
Combining Equations~(\ref{DenominatorResult}), (\ref{EasyNumeratorResult}), and (\ref{ComplicatedNumeratorResult}), we have
\begin{align}
A=& \frac{\det^2 \left( 1+ e^{X_{1,{\rm sp}}}e^{X_{0,{\rm sp}}}\right) }
{\det^2 \left( 1+ e^{X_{0,{\rm sp}}}\right)}  \braketop{j}{\frac{ e^{X_{1,{\rm sp}}}e^{X_{0,{\rm sp}}} e^{-X_{0,{\rm sp}}} }{1+e^{X_{1,{\rm sp}}}e^{X_{0,{\rm sp}}}}}{j} \nonumber \\
=& \det \left( \frac{   1+ e^{X_{1,{\rm sp}}}e^{X_{0,{\rm sp}}}  }{   1+ e^{X_{0,{\rm sp}}}  }  \right)^2
\braketop{j}{\frac{ e^{X_{1,{\rm sp}}} }{1+e^{X_{1,{\rm sp}}}e^{X_{0,{\rm sp}}}}}{j}
\end{align}

Now define $N \equiv e^{X_{0,{\rm sp}}}/(1+e^{X_{0,{\rm sp}}}), e^{X_{0,{\rm sp}}} = N/(1-N)$, the meaning of which is clear when $X_0 = -\beta H_0$.  Substituting $N$ for $e^{X_{0,{\rm sp}}}$, we obtain
\begin{align}
A=& \det \left( (1-N)+e^{X_{1,{\rm sp}}}N \right)^2 \left( \frac{N}{1-N}+e^{-X_{1,{\rm sp}}} \right)^{-1}_{jj} \nonumber \\
=& \det \left( (1-N)+e^{-itH_{1,{\rm sp}}}N \right)^2 \left( \frac{N}{1-N}+e^{itH_{1,{\rm sp}}} \right)^{-1}_{jj}.
\end{align}
We have reduced the expression to one involving operations on matrices in single-particle  Hilbert space.

\section{Effect of Core Hole in RSXS Spectra}
\begin{figure}[h]
\begin{center}
\includegraphics[width=0.8\linewidth,natwidth=557,natheight=377]{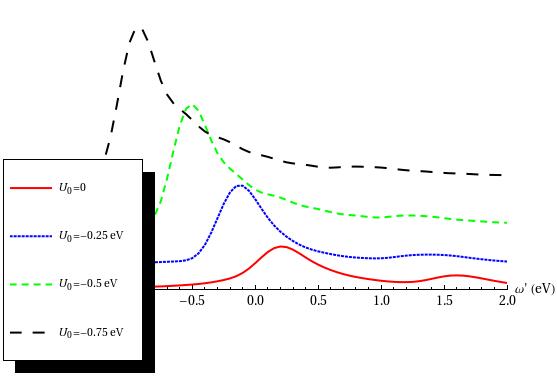} 
\end{center}
\caption{RSXS spectra of LBCO at different core hole strengths.  $\omega^\prime$ is defined as in Figure (1) of the main text.}
\label{CoreHoles}
\end{figure}

In Figure~(\ref{CoreHoles}) we show the effect of varying core hole strengths on the RSXS spectrum.  The essential features to note are (i) flattening of the high energy peak and growth of the low energy peak, and (ii) slight increase in peak separation, as core hole strength is increased.  By $U_0=-0.75$ eV, the upper peak is almost invisible.

The peak energies themselves depend on $U_0$ in that an attractive potential tends to lower all energies.  However, the separation between peaks depends only weakly on $U_0$.  Zero energy, i.e. the Fermi energy $E_f$, is not at the first maximum.  The location of this maximum is determined by dynamic nesting, which occurs at some energy above $E_f$, and the excitonic effect, which lowers energies of transient states.  Thus in principle the first peak could be at positive or negative energy.

\section{Core Hole Yukawa Potentials}
\begin{figure}[h]
\begin{center}
\includegraphics[width=0.8\linewidth,natwidth=459,natheight=311]{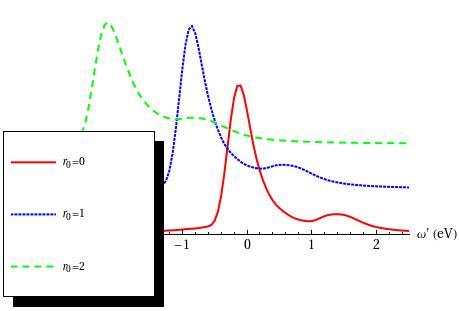} 
\end{center}
\caption{RSXS spectra of LBCO at different Yukawa potential ranges for fixed core hole potential depth $U_0=-0.25$ eV.  $\omega^\prime$ is defined as in Figure (1) of the main text.}
\label{Yukawa}
\end{figure}

Figure~(\ref{Yukawa}) shows the calculated RSXS spectra for parameters identical to those in the main text but Yukawa core hole potentials of the form
\begin{equation}
U(\vr) = U_0 \left\{ \begin{array}{cc} 1 & (\vr = {\bf 0}) \\
\frac{e^{-r/r_0}}{r} & (\vr \ne {\bf 0} \end{array} \right.,
\end{equation}
where $r_0$ is the range of the potential.  The effect is quite similar to that obtained by increasing $U_0$ in the case of a contact potential, which is sensible -- a potential well can become bigger by growing deeper or wider.  The high energy peak becomes less pronounced and both peaks move to lower energies.  For the Yukawa case, unlike for contact potentials, the separation between peaks decreases slightly as the range increases.  This is not profound -- it merely reflects the fact that different potentials yield different excitonic effects.  Incidentally, we note that it is possible that LDA-derived tight-binding parameters are quite accurate and that excitonic effects of a Yukawa potential reduce the separation of peaks to the experimentally-observed value.

\section{LDA- vs. ARPES-derived Band Structure}
\begin{figure}[h]
\begin{center}
\includegraphics[width=0.8\linewidth,natwidth=472,natheight=320]{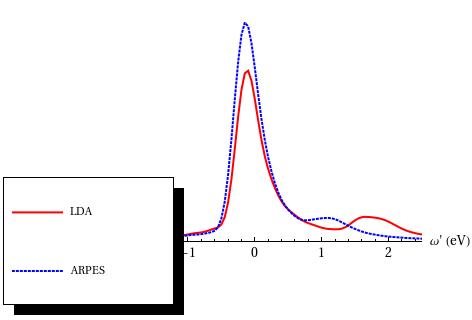} 
\end{center}
\caption{RSXS spectra of LBCO at different for LDA and ARPES tight-binding parameters.  $\omega^\prime$ is defined as in Figure (1) of the main text.}
\label{LDAvsARPES}
\end{figure}

Figure~(\ref{LDAvsARPES}) shows RSXS spectra using the same parameters as in the main text except for replacing hopping strengths with those obtained from LDA and ARPES.  The spectra are qualitatively the same but with different peak separations.  The experimentally-observed peak separation is less than that of LDA and more than that of ARPES.

\section{Slightly-Inelastic Scattering}
We assumed purely elastic scattering in which $\ket{f}=\ket{i}$.  However, resolutions of both energy and momentum are finite and, assuming $T=0$ for simplicity, a final state containing electron-hole pairs of very small total energy and momenta could correspond to apparently-elastic scattering.  We can give two simple reasons why this does not affect the validity of setting $\ket{f}=\ket{i}$.  Firstly, the rate of producing any given electron-hole pair, for example by scattering of Fermi sea electrons by the core hole potential, is independent of the transient state of the photoelectron.  Thus the creation of low-energy and low-momentum excitations in the final state will increase the measured elastic intensity uniformly by an $\ket{n}$-independent factor.  Secondly, matrix elements to produce a particle of momentum $\vq$ vary slowly as a function of $\vq$, hence slightly-inelastic processes do not exhibit a peak in momentum transfer.  They contribute a constant incoherent background which is easily isolated from the diffraction peak.  In practice, this is done by measuring RSXS spectra over a range of momentum transfer that crosses the ordering wavevector $\vQ$.